\documentclass[pre,aps,showpacs,twocolumn,floatfix,superscriptaddress]{revtex4}
\usepackage{epsfig}
\usepackage{color}
\usepackage{eqnarray,amsmath,amssymb}

\newcommand{\be}{\begin{equation}}
\newcommand{\ee}{\end{equation}}
\newcommand{\bea}{\begin{eqnarray}}
\newcommand{\eea}{\end{eqnarray}}
\newcommand{\ba}{\begin{array}}
\newcommand{\ea}{\end{array}}
\newcommand{\nn}{\nonumber}

\begin{document}
\title{Elastic traits of the extensible discrete wormlike chain model}

\author{Alessandro Fiasconaro}
\email{afiascon@unizar.es}
\affiliation{Departamento de F\'{\i}sica de la Materia Condensada,  Universidad de Zaragoza, 50009 Zaragoza,  Spain}
\affiliation{Instituto de Biocomputaci\'on y F\'{\i}sica de Sistemas  Complejos, Universidad de Zaragoza, Zaragoza, Spain}

\author{Fernando Falo}
\affiliation{Departamento de F\'{\i}sica de la  Materia Condensada, Universidad de Zaragoza, 50009 Zaragoza, Spain}
\affiliation{Instituto de Biocomputaci\'on y F\'{\i}sica de Sistemas  Complejos, Universidad de Zaragoza, Zaragoza, Spain}

\date{\today}
\begin{abstract}
Polymer models play the special role of elucidating the elementary features describing the physics of long molecules and become essential to interpret the measurements of their magnitudes. In this work the end-to-end distance of an extensible discrete worm-like chain polymer as a function of the applied force has been calculated both numerically and analytically, the latter as an effective approximation. The numerical evaluation uses the Transfer Matrix formalism to obtain an exact calculation of the partition function, while the analytic derivations generalize the simple phenomenological formulas largely used up to now. The obtained formulas are simple enough to be implemented in the fit analysis of experimental data of semi-flexible extensible polymers, with the result that the elastic parameters obtained are compatible with previous measurements, and more, their accuracy strongly improves  in a large range of chain extensibility.
\end{abstract}

\pacs{87.15.-v, 36.20.-r, 87.18.Tt, 83.10.Rs, 05.40.-a}
\keywords{Stochastic Modeling, Fluctuation phenomena, Polymer dynamics, Langevin equation}


\maketitle

\section{Introduction}
The technological advances in single molecule techniques (magnetic and optical tweezers, AFM, etc.) have allowed the manipulation and stretching of single polymer molecules by applying a longitudinal force and permitting the estimation of its elastic properties. In a celebrated experiment, Bustamante and collaborators have stretched a single double stranded DNA (dsDNA) molecule \cite{Busta1992} obtaining an extension curve as a function of a large interval of applied forces. 

Beyond the simplicity of this purely mechanical experiment, the precise modelling of its outcomes has occupied the energies of the researchers in a theoretical effort that still remains far to be completed.

The stretching features of a chain submerged in thermal fluctuations, have been described by means of the so-called worm-like chain model (WLC), which concerns a \emph{semi-flexible} continuous beam~\cite{1995Marko,1999Bouchiat}.
In one of the most common implementation of this idea, the WLC model has been discretized as a chain of connected rigid sticks with the inclusion of a transversal bending between them, so obtaining a discrete version of the WLC (DWLC) model~\cite{Rosa1,Rosa2,2004Lipo,2013Koslover,Manca2012JCP}. 
This model complicates the simpler freely jointed chain (FJC) model \cite{fjc}, composed by freely rotating rigid sticks which do not include any stiffness potential between links.  While the WLC models can describe the features of polymers presenting a bending elasticity, like the double stranded DNA (dsDNA), the FJC model can effectively depict the elastic features of a \emph{flexible} polymeric structure, whose paradigmatic example is the single stranded DNA (ssDNA)~\cite{Busta1992,storm2003} that presents a weak resistance to bend.

Moreover, the real polymers present a longitudinal elasticity that requires to add a new degree of freedom in both the FJC and the (D)WLC models.
To take into account the longitudinal extension, a simple correction has been introduced by Odijk \cite{odijk} in the WLC model by replacing the sticks with harmonic springs, and by adding phenomenologically the elastic contribution $f/(k l_0)$ to the end-to-end distance of the chain obtained with \emph{inextensible} bonds, where $f$ is the applied stretch force, $l_0$ the rest distance between consecutive monomers, and $k$ the longitudinal elastic constant of the links.

In this sense, the extension/force curve, normalized with the contour lenght of the chain,  is very simple 
\be
  \xi_N = \xi_{inext} + \frac{f}{k l_0},
  \label{naive}
\ee
or, in an alternative form  also largely used in the experimental literature,
\be
  \xi_M = \xi_{inext}  \left(1 + \frac{f}{k l_0} \right).
  \label{naiveL}
\ee
These expressions have been extensively used in fitting the experimental elastic properties of polymers as the ssDNA ~\cite{,storm2003,Grebikova2014,Grebikova2016Pol}, and the dsDNA \cite{Busta1996,Tskhovrebova1997,Rief1999,2002PRL_Bensimon,Wanga2001,Calderon2009,Frey2012,Bosco2014,Soler2016,2013JACSRicardo}, \emph{i.e.} with the FJC and the WLC models, respectively.

In recent works, a more complex expression related to the extensible FJC model has been derived~\cite{2009Balavaev,2019AF-FF,2020Buche,2022Buche}:
 \be
  \xi_{EF} = \mathcal{L}(\beta fl_0) + \frac{f}{kl_0} \left[1 + \frac{1-\mathcal{L}(\beta fl_0) \coth(\beta fl_0)}{1+\frac{f}{kl_0}\coth(\beta fl_0)} \right],
 \label{xi_E}
 \ee
with $ \mathcal{L}(x)=\coth{x}-1/x$ the Langevin function. Eq.~(\ref{xi_E}) permits an accurate analysis of the elastic parameters of a flexible polymer~\cite{2019AF-FF}.

Analogously, the bending elasticity of the molecular structures typical of the WLC model also requires an improved formulation with extensible links. Some of such studies have been done in Ref~\cite{2004Lipo,2013Koslover,Manca2012JCP}.

Purpose of this work is, on the one hand, to evaluate in an exact --numerical-- way the extensible discrete WLC model by means of the Transfer Matrix Theory. To do that, we perform an accurate calculation of the contribution to the partition function of the extensible degree of freedom, valid also for small chains. This allows to obtain numerically the extension curves for a wide range of the extension parameter. Not surprisingly, we find that, for highly extensible chains, the simple expressions of Eq.~(\ref{naive}) and Eq.~(\ref{naiveL}) clearly fails for low and intermediate forces. This indicates that such formulas are only a first order approach and a more elaborated analytical evaluation is needed. In this sense, and founding our calculations on the complete partition function, we are able, on the other hand, to write down some new handy formulas which nicely reproduce the exact Transfer Matrix calculations for almost all the force range here considered. These formulas can be easily implemented in the fit analysis of experimental data.

The paper is organized as follows: next section will present the details of the model and the calculation of the partition function; Section III will treat the numerical resolution in the outline of the Transfer Matrix Evaluation (TME); in Section IV we derive the analytic curves with the phenomenological generalization; Section V will present the fit procedure by using the approximated formula for the EDWLC model, by usign the TME data as reference; The Summary and Comments section will close the work.

\section{The model.}
\newcommand{\rad}{0.12cm}
\newcommand{\ds}{1.5cm}

\begin{figure}[h]
\centering
\includegraphics[angle=-0, width=0.85\columnwidth]{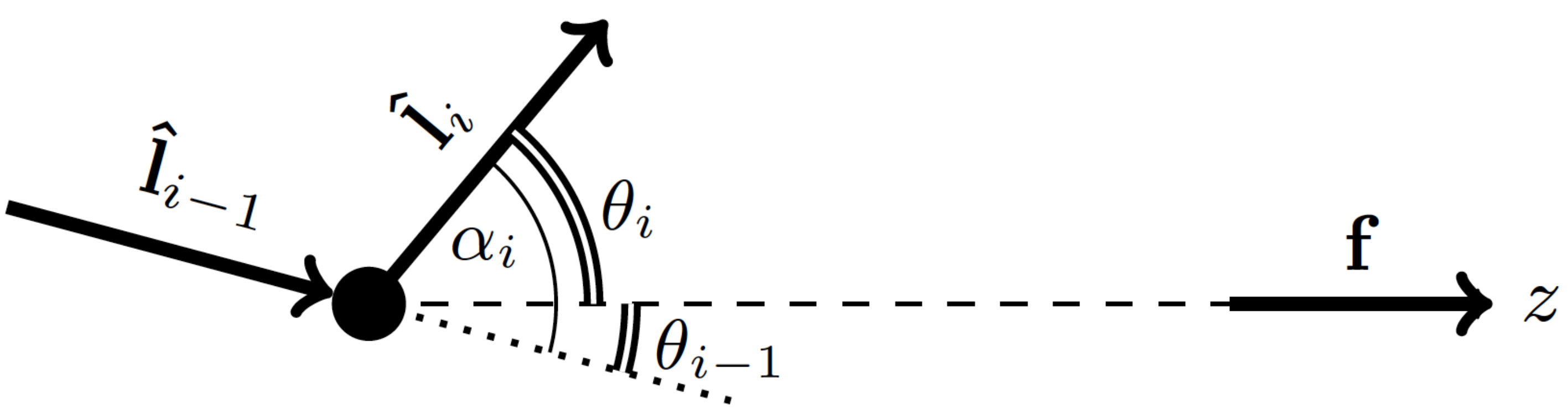}
\caption{Scheme of the bending recoil in the WLC model. The recoil torque energy of two consecutive links is $k_b \cos\alpha_{i,i-1}$, supposed the equilibrium angle $\theta_0=0$. The force $f$ pulls the chain along the $z$-axis. In the 3d space the three angles $\alpha_i$, $\theta_i$, and $\theta_{i-1}$ are not coplanar.}
\label{fig-WLCtheta}
\end{figure}

The Hamiltonian of the system is:
\be
 H = H_0+\sum_1^N -fl_i \cos\theta_i + \sum_1^N \frac{1}{2} k (l_i-l_0)^2 - k_b \sum_2^N \mathbf{\hat{l}}_i \cdot \mathbf{\hat{l}}_{i-1},
  \label{ham}
\ee
with $N$ the number of links, and $l_0$ the rest length of the springs, which corresponds to the Kuhn length of the polymer in the continuous case. $H_0=\sum_0^N p^2/2m$ is the kinetic energy contribution.  $\mathbf{\hat{l}}_i \cdot \mathbf{\hat{l}}_{i-i}$ is the scalar product between the unit vectors of two consecutive links, that is equal to $\cos\alpha_{i,i-1}$.  Specifically, $\mathbf{\hat{l}}_{i}=(\sin\theta_i \cos\phi_i,\sin\theta_i\sin\phi_i,\cos\theta_i)$, so 
\be
  \cos\alpha_{i,i-1}= \sin\theta_i \sin\theta_{i-1} \cos(\phi_{i}- \phi_{i-1}) +\cos\theta_i\cos\theta_{i-1}.
  \label{cos}
\ee
The partition function is then the sum over all the polymer configurations of $e^{-\beta H}$, specifically the spatial angles and spring length, written as follows:
\be
 Z =\sum_{\{\phi_i\}\{\theta_i\}\{l_i\}}e^{\beta \sum_{i=1}^N fl_i \cos\theta_i -\frac{1}{2} \beta k (l_i-l_0)^2+ \beta k_b \cos\alpha_{i,i-1}},
\ee
where the kinetic energy contributes with a force-independent multiplicative term, here omitted because not influential. In the last term, the sum starts from the index $i=2$, as the angle difference between the firsts links is not defined.

The partition function can be rewritten as follows:

\begin{widetext}
\begin{eqnarray}
 Z  &=&  \sum_{\{\phi_i\}\{\theta_i\}\{l_i\}} \prod_{i=1}^N e^{\beta fl_i \cos\theta_i-\frac{1}{2} \beta k (l_i-l_0)^2
                + \beta k_b  \cos\alpha_{i,i-1} } \nn \\
  &=& \int_{0}^{\pi} \cos\theta_1 d\theta_1 ... \cos\theta_N d\theta_N \int_{0}^{2\pi} d\phi_1 ... d\phi_N \int_{0}^{\infty} l_1^2 dl_1 ... l_N^2 dl_N \prod_{i=1}^N e^{\beta fl_i \cos\theta_i} e^{-\frac{1}{2} \beta k (l_i-l_0)^2 } \, e^{\beta k_b \cos\alpha_{i,i-1} } 
\end{eqnarray}
\end{widetext}
where the product and sum operators cannot be inverted, and the sum in all the possible configurations has been specified as the integral in the volume element $d\Omega=l^2\sin\theta \,dl \, d\phi \, d\theta $. The above expression is {\it not factorable} in single links because of the presence of the $\alpha_{i,i-1}$ term which involves the $\theta$ variable of two consecutive links: $i$ and $i-1$, while the term $\beta f \cos\theta_i$ affects the single links only (See Eq.~\ref{cos}). 

Nevertheless, both the integrals in the length variable $l$ and in the angular variable $\phi$ can be calculated for each link independently. The integral of the first variable is
 \be
G(\theta_i) =\int_{0}^{\infty} e^{\beta fl \cos\theta_i} e^{-\frac{1}{2} \beta k (l-l_0)^2 } l^2 dl,
 \label{1-int}
 \ee
and by changing variable $l-l_0 = x$, and approximating to high values of $kl_0$, so that $\beta k l_0^2 \gg 1$ we get  
\begin{widetext}
 \bea
 G(\theta_i) &=&e^{\beta fl_0 \cos\theta_i} \int_{-\infty}^{\infty}  e^{\beta f \cos\theta_i x - \frac{1}{2} \beta k x^2 } (x+l_0)^2 dx \nn \\
   &=&e^{\beta fl_0 \cos\theta_i} e^{\frac{\beta f^2 }{2 k}\cos^2\theta_i}  \sqrt{\frac{2\pi}{\beta^3 k^3}} \left[1+\beta kl_0^2\left(1 +  \frac{f}{ k l_0} \cos\theta_i \right)^2 \right] \nn \\
  &=& e^{\beta fl_0 \cos\theta_i} S(\theta_i). 
 \label{Gi}
 \eea
\end{widetext}

The integration limits of the above integral have been extended from $[-l_0,\infty]$ to $[-\infty,+\infty]$ after the change of variable.  
The added part of the integral, that involves the error function in the interval $[-\infty,-l_0]$, becomes negligible under the approximation at high $kl_0$, so allowing its analytic evaluation in Eq.~(\ref{Gi})~\footnote{The complete expression of $S(\theta_i)$ without the approximation $\beta kl_0^2 \gg 1$ is given by 
\begin{widetext}
$ S(\theta_i) = e^{\frac{\beta f^2 }{2 k}\cos^2\theta_i} \left\{ \sqrt{\frac{2\pi}{\beta^3 k^3}} \left[1 +\beta kl_0^2\left(1 +  \frac{f}{ k l_0} \cos\theta_i \right)^2 \right] \left[\frac{1}{2} +\frac{1}{2}erf(h) \right] + \frac{l_0}{\beta k} \left( 1 + \frac{f}{ k l_0} \cos\theta_i  \right) e^{-h^2} \right\}$ \\
with $h=\sqrt{\frac{\beta k l_0^2 }{2}} ( 1 + \frac{f}{ k l_0} \cos\theta_i)$ and $erf(\cdot)$ is the error function.
In our manuscript, the worst approximated condition is due for $k=10$ and $f=0$. With this choice, $h\approx 2.236$, $erf(h)\approx 0.9984$ and $e^{-h^2}\approx 6\cdot10^{-3}$, and the 2$^{\rm nd}$ addendum takes a value of the order of $6\cdot 10^{-4}$. So, the correction to $S(\theta_i)$ results negligible in all the presented cases. \end{widetext}}.

The term $S(\theta_i)$ represents the {\it extensible} contribution to the partition function. This expression is a higher approximation with respect to the analogous calculation presented by Kierfeld {\it et al.}~\cite{2004Lipo}, where the same contribution was expressed by the $e^{\frac{\beta f^2}{2k}\cos^2\theta}$ factor only, {\emph i.e.} the Hamiltonian there evaluated considers higher $k$-values than this one. At that level of approximation, the effective elastic contributions limits to the expression of Eq.~(\ref{naive}). In those conditions, they were able to resolve the integrals by using the spherical harmonics decomposition, that is not more helpful with the complete $S(\theta_i)$.

Concerning the $\phi$ integral, it is important to notice that the term $\cos(\phi_{i}- \phi_{i-1})$ included in Eq.~(\ref{cos}) concerns the freely rotating azimuthal angles of two subsequent links $\phi_i$ and $\phi_{i-1}$, whose difference ($\phi_{i}- \phi_{i-1}$) can be substituted by a singe variable $\phi$ in the corresponding integrations. In fact, given the periodicity of the cosine function in the $i$-th integral, the variable $\phi_{i-1}$ can be considered as a fixed phase that does not changes the integral evaluation. This gives
 \bea
 I_{\phi} (\theta_i,\theta_{i-1}) &=& \int_{0}^{2\pi} d\phi   e^{\beta k_b \cos\alpha_{i,i-1}}  \\
 &=&  e^{\beta k_b \cos\theta_i\cos\theta_{i-1}} \int_{0}^{2\pi} d\phi   e^{\beta k_b \sin\theta_i \sin\theta_{i-1} \cos\phi} \nn \\
 &=&  2 \pi e^{\beta k_b \cos\theta_i\cos\theta_{i-1}} I_0(\beta k_b \sin\theta_i \sin\theta_{i-1}) \nn
 \label{phi-int}
 \eea
 where $I_0(x)$ is the 0$^{\rm th}$-order Bessel function defined by $2\pi I_0(p)= \int_{0}^{2\pi} e^{p \cos\phi} d\phi$. 

As a result of the two above integrations, the partition function takes the shape:
%
\bea
 Z  =  \int_{-1}^{1} dx_1 &G(x_1) & \int_{-1}^{1} dx_2 ... dx_N G(x_2) I_{\phi}(x_2,x_1) ... \nn \\
     & &...  \,\,\,\,\,\,G(x_N) I_{\phi}(x_N,x_{N-1}) 
 \label{z2}            
\eea
where the change of variable $\cos\theta_i=x_i$ has been adopted.
Is is worth to note that in the limit of strong longitudinal stiffness the $S(\theta_i)$ function has to reach the value 1, \emph{i.e.} $\lim_{k\rightarrow\infty} S(\theta_i)=1$, which corresponds to the case of inextensible bonds.

As commented above, the expression (\ref{z2}) is evidently not computable at single link level. In order to calculate it numerically, we make use of the Transfer Matrix theory (see Ref.~\cite{Schneider1980}), which allows a precise determination of the force {\it vs} length curve.

\vspace{0.5truecm}
\section{Transfer matrix evaluation}

The result represented in Equation (\ref{z2}) can be written as
\bea
 Z  &=& \int_{-1}^{1} dx_1 G(x_1)  \int_{-1}^{1} dx_2 ... dx_i ...dx_N  \nn \\
     & & {\cal T}(x_2,x_1) ...  {\cal T}(x_i,x_{i-1}) ...  {\cal T}(x_N,x_{N-1})  
 \label{z3}            
\eea
with the general term ${\cal T}(x_i,x_{i-1}) =  G(x_i) I_{\phi}(x_i,x_{i-1})$.

The TFE consists in defining an integral operator of the kind:
\be
\int_{-1}^{1} dx \,{\cal T}(x',x) \psi_n(x) = \lambda_n \psi_n(x').
 \label{t1}            
\ee
with $\psi_n$ and $\lambda_n$ the $n$-th eigenfunction and the $n$-th eigenvalue respectively. 

If the integral (\ref{t1}) exists, the eigenfunctions are a basis which satisfy both the completeness and orthogonalization conditions: $\sum_n \psi^*_n(x) \psi_n(x') = \delta(x-x')$ and $\int dx \, \psi_n(x) \psi^*_{n'}(x)=\delta_{n,n'}$.

The integrals of Eq.~(\ref{z3}) are concatenated between each other by means of the eigenvalue equation~(\ref{t1}). Supposed $\psi_n(x)$ known, $G(x_1)$ can be decomposed as $G(x_1) = \sum_n a_n \psi_n(x_1)$, (with $a_n = \int dx G(x) \psi^*_n(x)$), and by substituting $G(x_1)$ in Eq.~(\ref{z3}) we get
\begin{widetext}
\bea
 Z  &=& \sum_n a_n  \int_{-1}^{1} dx_1 \psi_n(x_1) {\cal T}(x_2,x_1) \int_{-1}^{1} dx_2 ... dx_i ...dx_N   ...  {\cal T}(x_i,x_{i-1}) ...  {\cal T}(x_N,x_{N-1})  \nn \\
    &=& \sum_n a_n  \lambda_n \int_{-1}^{1} dx_2 \psi_n(x_2) {\cal T}(x_3,x_2) ... dx_i ...dx_N   ...  {\cal T}(x_i,x_{i-1}) ...  {\cal T}(x_N,x_{N-1})  \nn \\
    &=& \sum_n a_n  \lambda^{i-1}_n \int_{-1}^{1} dx_i \psi_n(x_{i-1}) {\cal T}(x_i,x_{i-1})  ...dx_N   ...    {\cal T}(x_N,x_{N-1})  \nn \\
    &=& ... \nn \\
    &=& \sum_n a_n  \lambda^{N-1}_n \int_{-1}^{1} dx_N \psi_n(x_N) = \sum_n a'_n  \lambda^{N-1}_n 
 \label{z4}            
\eea
\end{widetext}

The method results iterative, in the sense that the action of the operator transfer matrix (\ref{t1}) has the effect of shifting the index variable in each integration, getting one eigenvalue factor at each integration up to resolve all the chain, and finally obtaining the simple final expression of Eq.~(\ref{z4}). 
The eigenvalues can be put in decreasing order, then $\lambda_0 > \lambda_1 > ... $, and the greatest used as common factor:
\be
 Z = a'_0 \lambda^{N-1}_0 \left[ 1 + \sum_{n=1}\frac{a'_n}{a'_0}  \left( \frac{\lambda_n}{\lambda_0}\right)^{N-1} \right],
 \label{z5}
\ee
where the fractions $\frac{\lambda_n}{\lambda_0} <1$ for $n=1,...$.

The numerical evaluation of the previous expressions requires the discretization of the $x$-variable ($ -1 < x < 1$) in the integral (\ref{t1}), which results in the matrix eigenvalue equation
\be
 \sum_{k=0}^{N_a} {\cal T}(x_{ih},x_{jk}) \psi_n(x_{jk}) = \lambda_n \psi_n(x_{ih}),
 \label{t2}  
\ee
that can be diagonalized using standard methods in order to obtain the analogous of equation~(\ref{z5}) with a finite number of eigenvalues.

In the above expression, $N_a$ is the number of intervals ($\Delta x= 2/N_a$) in which the integral is discretized, and then $x_{ih} = -1 + h\Delta x, h= 0 ... N_a $. Along this work, we have used $N_a=4000$, that has guaranteed an optimum convergence of all the integrals calculated.

\vspace{0.2truecm}
\emph{Helmholtz function at the thermodynamic limit.---}

Equation (\ref{z5}) is the exact evaluation of the partition function in the case of $N$ links polymer chain. A simplified formula is obtained in the case of long chains, \emph{i.e.} at the thermodynamic limit.
\be
 \lim_{N \to \infty} Z = a'_0 \lambda^{N-1}_0.
 \label{z6}
\ee
In this case, the free energy $F=-k_{\rm B}T \ln Z$ can be calculated as 
\bea
 F &=&  -k_{\rm B}T  \ln \lambda^{N-1}_0 -k_{\rm B}T \ln a'_0 \nn \\
   &=& -k_{\rm B}T N \ln \lambda_0  -k_{\rm B}T \ln \frac{a'_0}{\lambda_0},
 \label{f1}
\eea
and, again in the thermodynamic limit, the last constant term can be neglected with respect to the first one that scales as $N$, so that
\be
 F \approx  -k_{\rm B}T N \ln \lambda_0. 
 \label{f2}
\ee

\vspace{0.2truecm}
\emph{End-to-end distance.---}
Given the Helmholtz function $F$, the normalized end-to-end distance along the direction of the force is given by the average:
 \be
  \xi = \frac{1}{Nl_0} \langle l \cos\theta\rangle = -\frac{1}{Nl_0} \frac{dF}{df}.
 \label{dee1}
 \ee

The numerical evaluation of $\xi$ with the Transfer Matrix methods, that we call here $\xi_{\rm TME}$, consists in discretizing Eq.~(\ref{t1}) and construct the ${\cal T}(x_i,x_j)$ matrix for a certain value of force $f$, then diagonalize it by taking the highest eigenvalue $\lambda_0$, and finally derive numerically applying Eq.~(\ref{dee1}), according to the chosen $f$-span. The results have been shown in Fig.~\ref{TME-LE}, together with the dynamical simulations of the Langevin equation which have been computed to compare and double-check the TM results.

\subsection{Symmetrical Transfer Matrix Algorithm.---}

The expression~(\ref{z2}) can be written in an explicit symmetrical form as follows:
\begin{widetext}
\be
 Z  \!=\!  \int_{-1}^{1} dx_1 G^{1/2}(x_1) \int_{-1}^{1} dx_2 ... dx_N  G^{1/2}(x_1)  I_{\phi}(x_2,x_1) G^{1/2}(x_2)...  G^{1/2}(x_{N-1}) I_{\phi}(x_N,x_{N-1}) G^{1/2}(x_N) G^{1/2}(x_N).            
\ee
\end{widetext}
In this case, the general integrand ${\cal T}(x_i,x_{i-1})=G^{1/2}(x_i)I_{\phi}(x_i,x_{i-1}) G^{1/2}(x_{i-1})$ is symmetrical with respect to the change $i$ and $i-1$. This way the expression (\ref{z3}) becomes

\bea
 Z  &=& \int_{-1}^{1} dx_1 G^{1/2}(x_1)  \int_{-1}^{1} dx_2 ... dx_i ...dx_N \\
     & & {\cal T}(x_2,x_1) ...  {\cal T}(x_i,x_{i-1}) ...  {\cal T}(x_N,x_{N-1}) G^{1/2}(x_{N}).   \nn 
 \label{z3s}            
\eea

By discretizing the integrals in a similar way as Eqs.~(\ref{t1}) it is possible to write down an eigenvalue equation similar to Eq.~(\ref{t2}). In this symmetric case, the eigenfunctions are necessarily real $\psi_n^*(x)=\psi_n(x)$. This means that $\sum_n \psi_n(x) \psi_n(x') = \delta(x-x')$ and $\int dx \, \psi_n(x) \psi_{n'}(x)=\delta_{n,n'}$.
By supposing the $\psi_n(x)$ eigenfunctions known, we can write down the first integral of Eq.~(\ref{z3s}) by decomposing $ G^{1/2}(x_1) = \sum_n a_n \psi_n(x_1)$, (with $a_n = \int dx G^{1/2}(x) \psi_n(x)$), and substituting in Eq.~(\ref{z3s}) we get, equivalently as Eq.~(\ref{z4}), and with identical procedure

\begin{widetext}
\bea
 Z  &=& \sum_n a_n  \int_{-1}^{1} dx_1 \psi_n(x_1) {\cal T}(x_2,x_1) \int_{-1}^{1} dx_2 ... dx_i ...dx_N   ...  {\cal T}(x_i,x_{i-1}) ...  {\cal T}(x_N,x_{N-1}) G^{1/2}(x_N)  \nn \\
    &=& ... \nn \\
    &=& \sum_n a_n  \lambda^{N-1}_n \int_{-1}^{1} dx_N \psi_n(x_N) G^{1/2}(x_N) = \sum_n a_n^2  \lambda^{N-1}_n.
 \label{z4s}            
\eea
\end{widetext}
Along this work, given the simplification of the calculations provided by the real eigenvalues, the symmetrical description has been adopted.
The practical implementation follows the same procedure as described after Eq.~(\ref{z4}). 

\subsection{Langevin simulations.---}
In order to check the numerical result of transfer matrix we have performed Langevin dynamical computer simulations of the DWLC. The polymer simulated consists of $N+1$ dimensionless monomers connected by harmonic springs, interacting with a bending energy (the total potential energy is indicated with $V(\mathbf{r})$), pulled by a  constant force $f$ in the $z$-direction in accordance to the Hamiltonian (\ref{ham}).

The dynamics of the chain is given by the overdamped Langevin equation (LE) of motion
 \be
 \dot{\mathbf{r}_i} = - \vec{\nabla}_i V(\mathbf{r}_i) + \mathbf{f}\cdot\mathbf{\hat{z}} \delta_{i,N} + \sqrt{2k_BT} \vec{\eta}_i(t),
 \label{eq}
 \ee
where $\vec{\eta}_i(t)$ represents the thermal contribution in the shape of a Gaussian uncorrelated noise: $\langle\vec{\eta}_{i,\alpha}(t)\rangle=0$, and $\langle\vec{\eta}_{i,\alpha}(t)\vec{\eta}_{j,\beta}(t')\rangle=\delta_{i,j}\delta_{\alpha,\beta}\delta(t-t')$, where $i,j=1,...,(N+1)$, $\alpha, \beta=x,y,z$. The nabla operator is defined as $\vec{\nabla}_i = \partial / \partial x_i \mathbf{i} + \partial / \partial y_i \mathbf{j}  + \partial / \partial z_i \mathbf{k}$.
The constant force $f$ pulls the last monomer in order to stretch dynamically the polymer, while the first monomer is held fixed. 
The simulations have been performed by averaging the \emph{end-to-end} distance in a long trajectory by integrating Eq.~(\ref{eq}) with a 2$^{\rm nd}$ order Runge-Kutta algorithm~\cite{RK} by using an integration time step $\Delta t=0.001$.

\begin{figure}[tb]
\centering
\includegraphics[angle=-0, width=1.0\columnwidth]{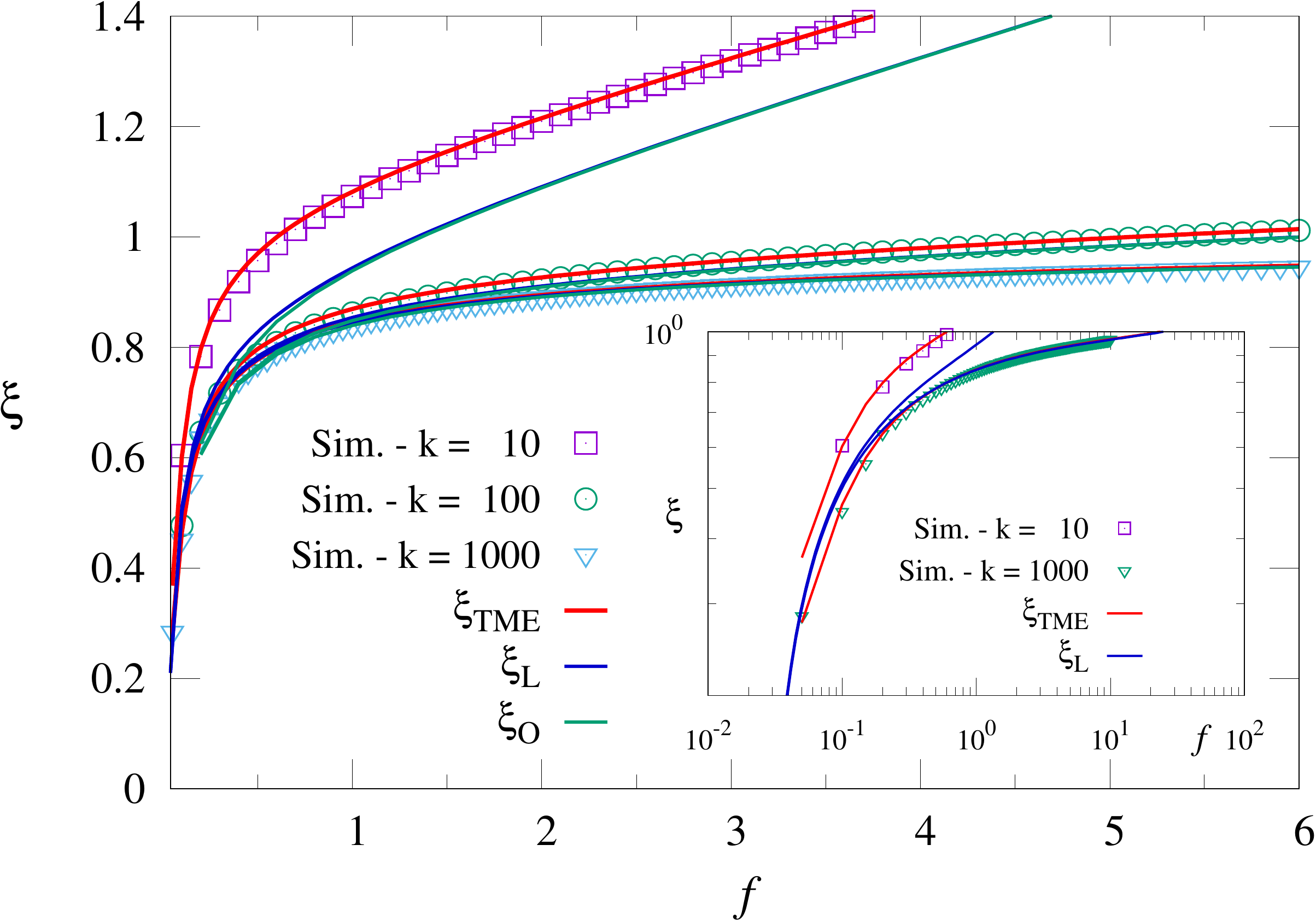}
\caption{Normalized end-to-end distance $\xi$ as a function of the force $f=$ the extensible discrete WLC model, for different elastic constant $k$, with the bending constant $k_b=10$ and $l_0=1$. The symbols represents the Langevin simulations, the curves which superimpose with the symbols are the transfer matrix evaluation. $\xi_L$ and $\xi_O$ (Eq.~(\ref{xi_L}) and Eq.~(\ref{xi_O})) are the \emph{na\"ive} extensible generalization (with $f/kl_0$) to the discrete inextensible WLC formulas by Rosa \emph{et al.} at high forces (Eq.~(\ref{RosaHF})) and all forces interpolation (Eq.~(\ref{WR})), respectively.}
\label{TME-LE}
\end{figure}

The comparison between the TME and the LE are shown in Figs.~\ref{TME-LE} and ~\ref{TME-N}, where we find an excellent agreement between them, which confirms the correctness of the calculations performed with the transfer matrix methods.
The parameters used in these calculations are: the bending constant $k_b=10$, the rest length $l_0=1$, and $\beta=1$, which are the standard magnitude values used along the manuscript, and the longitudinal elastic constant $k = 1000$, that will be changed in some of the calculation performed.
Since the TM calculations are much faster than the Langevin simulations, we used them to provide the statistical data on which to perform the fit analysis with the analytical formulas found in the next section.

\subsection{Dependence on N.---}

\begin{figure}[tbp]
\centering
\includegraphics[angle=-0, width=1.0\columnwidth]{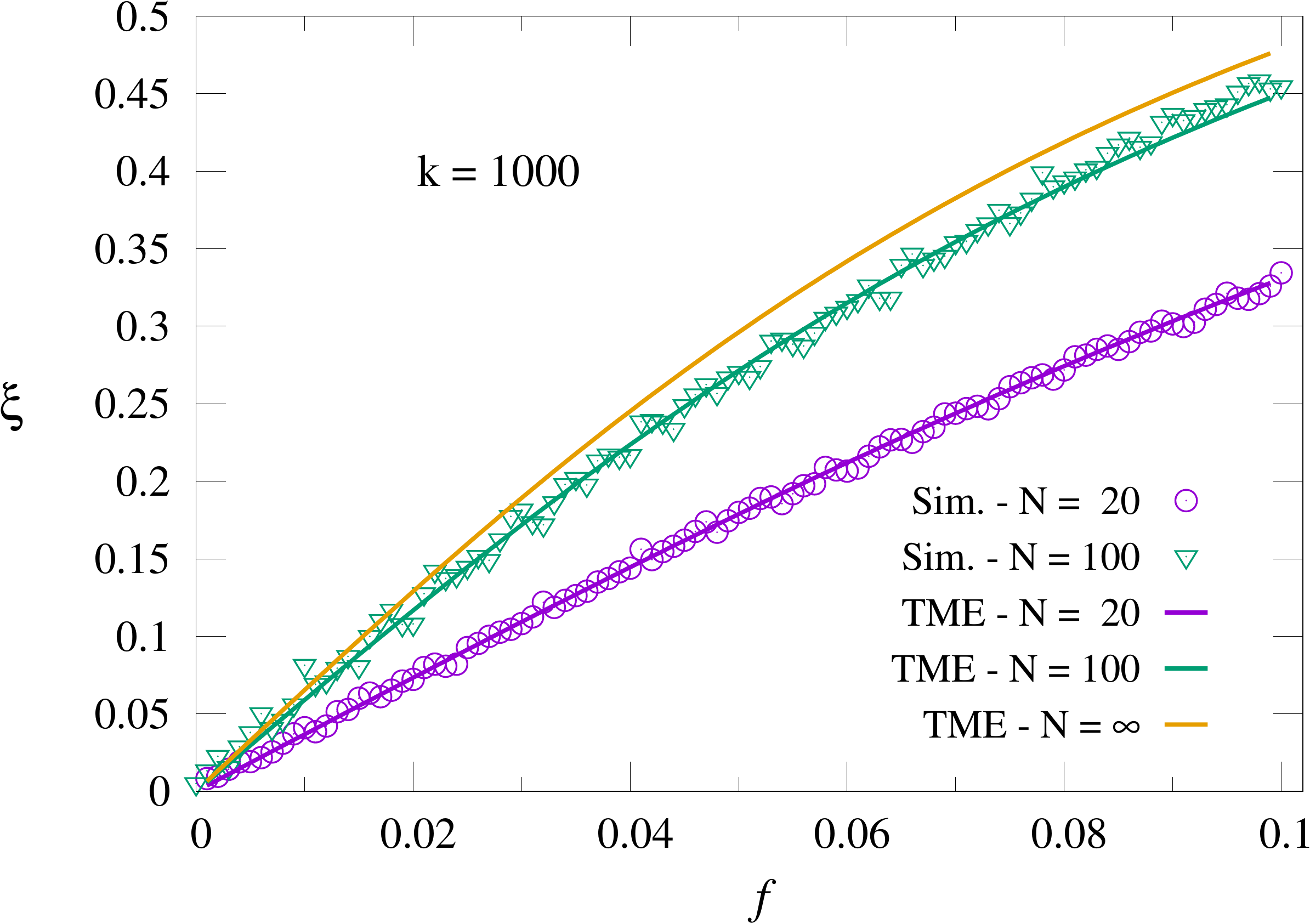}
\caption{The curves of the normalized end-to-end distance $\xi$ as a function of the force $f$ for the extensible WLC model calculated with the Transfer Matrix method at different polymer lengths, specifically $N=20$ and $N=100$, compared with the thermodynamic limit $N=\infty$. The longitudinal elastic constant is $k = 1000$, the bending constant is $k_b=10$, the rest length $l_0=1$, and $\beta=1$. The symbols represent the Langevin simulations with the same parameters.}
\label{TME-N}
\end{figure}

The TM numerical evaluation also allows to calculate the dependence of the end-to-end distance with the number of links $N$ by directly using the partition function of Eq.~(\ref{z5}). In this case a larger number of eigenvalues have to be taken into account, instead of the largest one only (in principle all $N_a$ values, but in practice the first four are generally enough), and in addition it is necessary to evaluate all the corresponding integrals $\int_{-1}^{1} dx \psi_n(x)$ and $\int_{-1}^{1} dx \psi_n(x) G(x) $ of Eq.~(\ref{z4}) (or $\int_{-1}^{1} dx \psi_n(x)G^{1/2}(x)$ of Eq.~(\ref{z4s}) in the symmetric case). The comparison for different polymer lengths with the result at the thermodynamic limit is presented in Fig.~\ref{TME-N}. 
The results have been shown at relatively low forces, where the size effects are more evident.

\section{Analytic approximations of extensible links.}

\subsection{Phenomenological extensible links.}
The \emph{inextensible} discrete WLC model has been studied by Rosa \emph{et al.}~\cite{Rosa1,Rosa2}, who have been able to write down the following analytic expressions that interpolates all the force range they considered: 
 \bea
  \beta f l_0 &=&2\beta k_b \left[ \sqrt{ 1 + \frac{1}{(2\beta k_b)^2} \frac{1}{(1-\xi)^2} } -  \sqrt{ 1 + \frac{1}{(2\beta k_b)^2}} \right] \nn \\ 
            &+ & \left( 3\frac{1-{\cal L}(\beta k_b)} {1+{\cal L}(\beta k_b)} - \frac{1/(2\beta k_b)}{\sqrt{1+1/(2\beta k_b)^2}} \right) \xi,
 \label{WR}
 \eea
with ${\cal L}(x)=\coth x-1/x$, which tends to the continuous case for $\lim_{l_0 \to 0}$ giving the famous Marko and Siggia equation $\beta L_P f = \frac{1}{4(1-\xi)^2} - \frac{1}{4}+\xi$, with $\beta k_b l_0=L_P$. 

Another --much simpler-- expression is the \emph{high force} approximation 
 \be
 \xi_{H} =1-\frac{1}{\beta \sqrt{ (l_0 f)^2 + 4 l_0 k_b f }},
 \label{RosaHF}
 \ee
which differs from Eq.~(\ref{WR}) only at very low forces, and it is here considered as a good reference. 
As commented in the introduction, these formulas can be naively generalized for the extensible polymers by adding the elastic contribution $f/(kl_0)$ to every single link, procedure that is largely used in the literature for fit purposes. Specifically, the high force limit gives the handy expression
 \be
  \xi_{L} = \xi_{H}  +  \frac{f}{k l_0}.
  \label{xi_L}
 \ee
while Eq.~(\ref{WR}) generalizes to the extensible case by changing 
 \be 
   \xi \to \xi_O = \xi - \frac{f}{kl_0},
 \label{xi_O}
 \ee
whose inversion to provide a function $\xi(f)$ is not analytically feasible, and for this reason only the high force approximation Eq.~(\ref{xi_L}) is useful in fitting analysis.

\subsection{Analytic approximations.}
In order to provide a simple expression able to improve the na\"ive formula~(\ref{xi_L}), we notice that if we consider constant the value of the cosine present in function $S(\theta_i)$ of Eq.~(\ref{Gi}), its contribution to the partition function can be factored with respect to the rest of integrands in Eq.~(\ref{z2}), leaving the partition function of the \emph{inextensible} discrete WLC as a global factor. This way our approximated proposal is to write the partition function as follows:
\be
 Z  \approx S^N(\tilde{\theta}) Z_{W},
 \label{z10}            
\ee
where $S(\tilde{\theta}) = e^{\frac{\beta f^2}{2k}\cos^2\tilde{\theta}}  \sqrt{\frac{2\pi}{\beta^3 k^3}} \left[1+\beta kl_0^2(1 + \frac{f}{k l_0}\cos\tilde{\theta} )^2 \right]$.
In other words, the $S$ function, approximated as constant as concerns the cosine term at every force values, contributes to the partition function as an external factor that accounts for the longitudinal elasticity, while the 2$^{\rm nd}$ factor $Z_{W}$ continues being the contribution of the discrete version of the worm-like chain model that includes the bending degree of freedom of the polymer.

\vspace{0.2truecm}
\emph{End-to-end distance.---}
The normalized end-to-end distance along the direction of the force is again given by:
 \be
  \xi = \frac{\langle l \cos\theta\rangle}{l_0}  = -\frac{1}{Nl_0} \frac{dF}{df} = \frac{1}{N \beta l_0 Z } \frac{dZ}{df},
 \label{dee}
 \ee
which allows to evaluate the expression 
 \bea
  \xi_{EW} &=& \frac{1}{N\beta l_0 S(\tilde{\theta}) Z_{W}}  S(\tilde{\theta}) \frac{dZ_W}{df} +   \frac{1}{\beta l_0 S(\tilde{\theta}) Z_{W}}  \frac{dS(\tilde{\theta})}{df}Z_W \nn \\
  &=& \xi_{W}  +  \frac{1}{\beta l_0 S(\tilde{\theta})}  \frac{dS(\tilde{\theta})}{df} \nn \\
  &=& \xi_{W}  +  \frac{f\cos^2\tilde{\theta}}{k l_0} +  \frac{ 2\cos\tilde{\theta} (1+\frac{f}{kl_0}\cos\tilde{\theta})  }{1+\beta k l_0^2 (1+\frac{f}{kl_0}\cos\tilde{\theta})^2 },
  \label{xi_EW}
 \eea
with $\xi_W$ the normalized end-to-end distance of the discrete WLC, supposed known.

In order to write down a clear handful formula it is necessary to characterize the  $\cos\tilde{\theta}$ term. As a zero-th order approach, we can substitute in the above expression the value $\cos\tilde{\theta}\approx 1$, that make sense at very high force values, by obtaining the simple expression
 \be
  \xi_{EW_0} \approx  \xi_W  +  \frac{f  }{k l_0} +  \frac{ 2 (1+\frac{f}{kl_0})  }{1+\beta k l_0^2 (1+\frac{f}{kl_0})^2 }.
  \label{xi_EW0}
 \ee
 The two first terms in the above expression are nothing but the na\"ive correction to the extensible chain commented in the introduction (Eq.~(\ref{naive})).
 
An improved approximation can be obtained by substituting the cosine with the $\xi_W$ expression.
In fact $\xi_W$ is, essentially, the projection of the unit vector of the single polymer link along the $f$-direction in the inextensible chain, or, in other words, it is the suitable $\cos\tilde{\theta}$ average of the polymer  links. This way $\xi_W$ can be used to approximate the cosine in the end-to-end expression~(\ref{xi_EW}). 
 That considered we obtain
 \be
  \xi_{EW_1} \approx  \xi_W  +  \frac{f  }{k l_0}\xi_W^2 +  \frac{ 2 \xi_W (1+\frac{f}{kl_0}\xi_W)  }{1+\beta k l_0^2 (1+\frac{f}{kl_0}\xi_W)^2 },
  \label{xi_EW1}
 \ee
which represents a better approximation than the previous ($\cos\tilde{\theta}\approx1$), being  $\xi_W=\xi_W(f)$ a function that depends of the applied force. 

Along this latter line of reasoning, a third possible proposal arises by the again considering $\cos\tilde{\theta}\approx \xi_W$, but substituting it in $S(\tilde{\theta})$ instead than in Eq.~(\ref{xi_EW}). This way we have  
\be
 S(\xi_W) = e^{\frac{\beta f^2}{2k} \xi_W^2}  \sqrt{\frac{2\pi}{\beta^3 k^3}} \left[1+\beta kl_0^2 (1+ \frac{f}{kl_0} \xi_W )^2 \right],
 \label{Sq}  
\ee
which derived according to equation~(\ref{dee}) generates an additional term because of the $f$-dependence of $\xi_W$, so obtaining
\begin{widetext}
\bea
  \xi_{EW_2} &=& \frac{1}{N\beta l_0 S(\xi_W) Z_{W}}  S(\xi_W) \frac{dZ_W}{df} +   \frac{1}{\beta l_0 S(\xi_W) Z_{W}}  \frac{dS(\xi_W)}{df}Z_W = \xi_W  +  \frac{1}{\beta l_0 S(\xi_W)}  \frac{dS(\xi_W)}{df} \nn \\
  &=& \xi_W  + \left[ \frac{f}{k l_0}\xi_W +  \frac{ 2(1+\frac{f}{kl_0}\xi_W)  } {1+\beta k l_0^2 (1+\frac{f}{kl_0}\xi_W)^2 } \right] \left( \xi_W + f\frac{d\xi_W}{df} \right).
  \label{xi_EW2}
\eea
\end{widetext}

Given the discrete nature of this model, the best expression for $\xi_W$ is provided by Eq.~(\ref{RosaHF}) calculated for the inextensible discrete WLC model in Ref.~\cite{Rosa1,Rosa2}. The explicit expression are then obtained with the substitution:
\be
 \xi_W \rightarrow \xi_H.
\ee
All the above approximations are in principle valid at high forces. However, the discrepancies of the TM evaluations and the analytic curves are evident only at very low forces $f\ll1$, and for this reason the lower extreme if the fit analysis can be considered equal to zero without big loss of precision. 

Beside the fact that the three equations~(\ref{xi_EW0}), (\ref{xi_EW1}) and (\ref{xi_EW2}) present, in the order, an increasing complexity, they are still formed by elementary functions, and are suitable to be easily inserted in any fitting tool to analyze experimental data.

\begin{figure}[tbp]
\centering
\includegraphics[angle=-0, width=1.0\columnwidth]{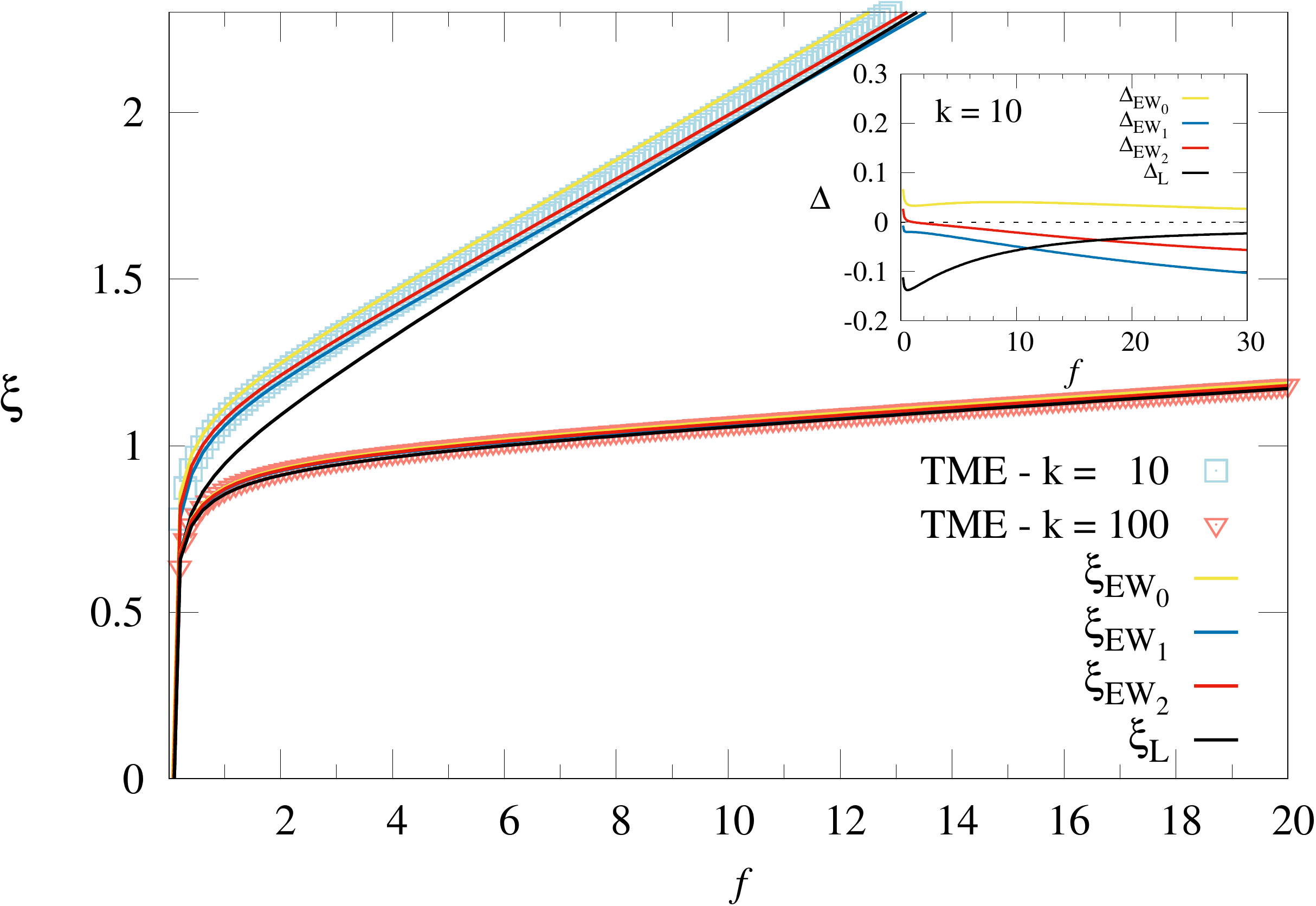}
\caption{Comparison of the approximated formulas (lines) with the ``exact" TME (symbols) with $k=10,100$.}
\label{EDWLC_teor}
\end{figure}

Figure~\ref{EDWLC_teor} shows the above calculated formulas and the na\"ive equation~(\ref{xi_L}), compared with the TM evaluation for two elastic constants: $k=10$ and $k=100$. We can see that the curves reveal their differences only at low $k$ values (for $k=1000$ they completely overlap a this scale). The inset of the figure shows the difference of the functions with respect to the TME. We can observe that the na\"ive approximation clearly deviates form TME at low and moderate forces and thus it is unable to predict the entropic region of the curve dominated by the bending energy term. However, the formulas $\xi_{EW_0}$ $\xi_{EW_1}$ and $\xi_{EW_2}$ nicely follow the TME curve, and for this reason it is expected that they give a good parameter prediction in the fit that includes the entropic region. Strangely enough, we can see how at high forces the formulas $\xi_{EW_1}$ and $\xi_{EW_2}$ do not converge to the TME curve and the difference increases with $f$. On the contrary, the expression $\xi_{EW_0}$ and the na\"ive $\xi_L$ present the correct trend for $f\rightarrow \infty$. 

Despite this behavior, due to the cosine square term that multiplies $f/(kl_0)$ in the 2$^{nd}$ term of Eq.~(\ref{xi_EW1}) and the equivalent one in Eq.~(\ref{xi_EW2}), $\xi_{EW_2}$ generates the best parameter predictions with respect to both $\xi_{EW_0}$ and $\xi_{EW_1}$ when used in the fit analysis.

The general result is that the expression (\ref{xi_EW2}) gives the best performances for low values of the longitudinal elastic parameter $k$, region in which the deviation of the $d_{ee}$ from the inextensible case becomes more evident. Also, it gives the best evaluation at intermediate $k$ values, provided that the contour length of the chain can be considered known. The fit analysis is the subject of next section.

\section{Fit results}
In order to check the goodness of the analytic formulas calculated in the previous section $\xi_{EW_0}$, $\xi_{EW_1}$, $\xi_{EW_2}$, and their ability to estimate the parameters of the model, we have performed the fit on the TME data --used as reference-- by using those formulas, together with the extensible modification of the Rosa formula ($\xi_L$) suitable in this model more than that by Marko and Siggia because of the discrete nature of the model. The TME provides the best values for this purpose because the data are essentially ``exact", as they come out of a numerical evaluation of the complete partition function.

The fit parameters that appear in the model are the longitudinal elastic constant $k$, the bending constant $k_b$, and the rest length of the links $l_0$. Moreover, to be realistic, the functions need to be defined with an additional parameter that takes into account the polymer length. In other terms, the fitted expression is considered of the type
\be
 \xi(f;k,k_b,l_0,L_c) \rightarrow L_c \xi(f;k,k_b,l_0)
\ee
where the parameter $L_c$ is the contour length of the polymer that multiplies the normalized extension.

By comparison, the reported fits includes those obtained by using both the extensible FJC model ($\xi_{EF}$) and the linear function typically used in the high force region of the curve, \emph{i.e.} $f>30$ in our choice.

Concerning the linear behavior, it is worth to underline that its expression can be written as $\xi=L_c(1+\frac{f}{S})$, where the $S$ parameter represents the product $kl_0$. So the two latter constants, which are parameters of the model, result indistinguishable between each other in the linear fit, though $S$ is a magnitude often obtained from either the fit of experimental measures \cite{2013JACSRicardo} or theoretical studies \cite{2019PRLPerez}.

According to their derivation, also the formulas $\xi_L$ and $\xi_{EW_0}$ have been fitted at high forces only, while $\xi_{EW_1}$ and $\xi_{EW_2}$ have been applied to all the range calculated with the TM methods: $f\in \, ]0,50]$, so including the low force region, simplifying of the fit procedure.

The results of the parameters prediction are resumed in Table~\ref{table4-4m}, calculated for three bond elastic constants (namely $k=10$, 100, and 1000) in a four parameters' fit. The reference data TME for the fit consist of 500 points generated with a force step $\Delta f =0.1$ and an integral discretization $N_a = 4000$. 
We can observe how the best outcomes at high elastic constant ($k=1000$) are still given by the Rosa formula $\xi_L$, which provides the best evaluation of the parameters, closely followed by the $\xi_{EW_0}$. In this $k$ region, the longitudinal elasticity becomes less relevant and the inextensible approximation reveals to give the expected parameter values. For the lower elasticity value $k=100$, $\xi_L$ already provides very bad parameters' predictions and the first approximated formula $\xi_{EW_0}$ gives the best results. For low elastic values $k=10$, $\xi_{EW_2}$ (formula~(\ref{xi_EW2})) gives the best predictions.

\begin{table}[tbp]
\begin{center}
\begin{tabular}{c c c c c c | c c }
\hline \hline
                       &             & $\xi_{EW_0}$   & $\xi_{EW_1}$   &  $\xi_{EW_2}$   &$\xi_{L}$    & $\xi_{EF}$      & $L_c(1\!+\!f/S)$  \\
                       &             &  $f\!>\!30$            &   $]0,50]$   &   $]0,50]$    &  $f\!>\!30$    &  $f\!>\!30$          &  $f\!>\!30$       \\
\hline
                         & $k$        &  6.37            &  6.62              & {\bf 7.98}       &  1.73       &  1.70           & ({\bf 10.00})  \\ 
$k=10$             & $k_b$    &  3.08              &  7.15              &  {\bf 7.77}      &  30.64     &  -                 &  -    \\
                         & $l_0$     &  1.538            &  1.582            & {\bf 1.281}     &  5.837     &  5.902         &  -      \\
                         & $L_c$    &  0.982            &  1.057            & 1.029            &  1.007     &  1.003         & {\bf 1.000}  \\
\hline
                          & $k$        & {\bf 91.40}    & 136.6               & 168.0        &  58.63     & 43.96        &  ({\bf  95.40})    \\
$k=100$           & $k_b$     & {\bf 10.50}    & 12.48               & 15.00        &  7.97       & -                &  -   \\
                         & $l_0$      & {\bf 1.080}    & 0.718               & 0.580        & 1.709      & 2.296       &  -   \\
                         & $L_c$     & 0.992           & {\bf 1.000}        & 0.993        & 1.002      &  0.998       &  0.979 \\

\hline
                            & $k$         & {\bf 947.6}      & 5391      & 5674           & {\bf 946.6}     & 386.2        &  (725.0)\\
$k\!=\!1000$            & $k_b$     & {\bf 9.47}        & 49.20     & 50.88          & {\bf 9.47}       &  -               &  -       \\
                            & $l_0$      & {\bf 1.051}      & 0.179     & 0.171          & {\bf 1.055}     &  2.396       &  -         \\
                            & $L_c$     & {\bf 0.999}      & 1.000     & 0.999          & {\bf 1.001}     &  0.991       &  0.970 \\

\end{tabular}
\end{center}
\caption{Four parameters fit ($k$, $k_b$, $l_0$, $L_c$) by using the analytic approximation Eq.~(\ref{xi_EW0}), Eq.~(\ref{xi_EW1}) and Eq.~(\ref{xi_EW2}), and the na\"ive formula $\xi_L$ of Eq.~(\ref{xi_L}). We have used the reference values: $l_0=1$, and $k_b=10$, and three values of the elastic constant: $k=10$, $k=100$ and $k=1000$. The linear fit, included here for completeness, is only able to provide the $L_c$ parameter, as the elastic parameter $k$ is included together with the length $l_0$ in the ``force" term $S=kl_0$. The bolded values indicate the best fit outcomes. In parenthesis the $S$-value of the linear fit, which coincide with $k$ in this example.}
\label{table4-4m}
\end{table}

 \begin{figure}[bp]
\centering
\includegraphics[angle=-0, width=1.0\columnwidth]{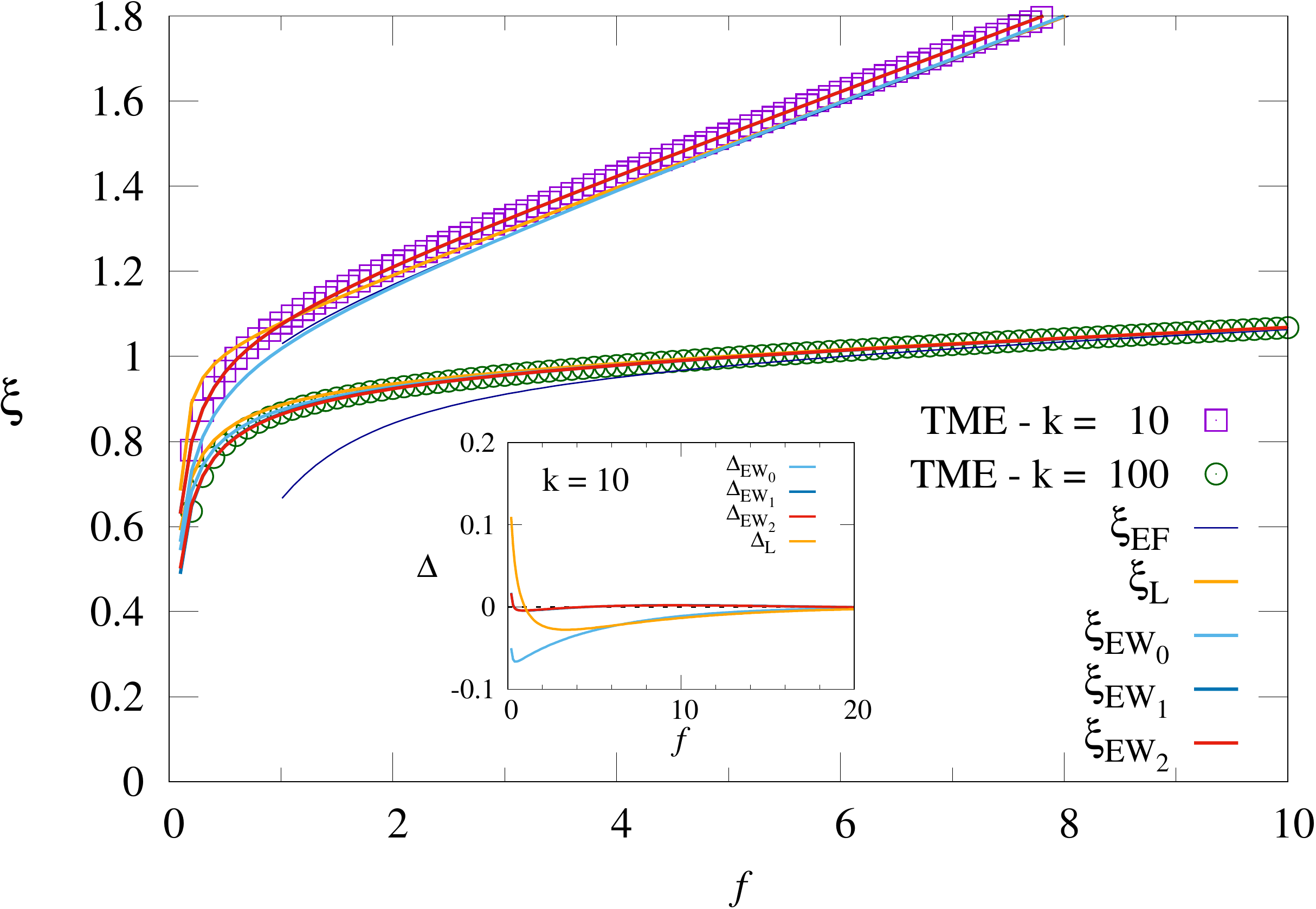}
\caption{Fit of the TME data, by using the analytical formulas discussed in the text, by using four free parameters $k$, $k_b$, $l_0$ and $L_c$. The values obtained are reported in Table~\ref{table4-4m}.}
\label{EDWLC_fit}
\end{figure}
In Fig.~\ref{EDWLC_fit} the curve calculated with the TME has been plotted for two different values of the elastic constant $k=10$ and $k=100$, together with the fitting curves. 
It is possible to see that all the curves practically overlap for $k=100$, and even more they overlap for $k=1000$ (not shown), while they clearly distinguish for low $k$s, especially at low forces. 
The inset of the figure reports the differences of the fitting functions with the TME, showing that $\xi_{EW_1}$ and $\xi_{EW_2}$ perfectly reproduce the TME curve in all the force range, while $\xi_L$ and $\xi_{EW_0}$ clearly fails at low and intermediate forces.

Generally, an increase of the performance in the fit is achieved by reducing the number of free parameters. The most delicate parameter here is the rest length of the individual links $l_0$, which is able to significantly change the prediction of both the elastic constants $k$ and $k_b$. Unfortunately, $l_0$ is also the most difficult parameter to obtain and the fit would strongly profit of its independent evaluation. A good alternative relies in the contour length $L_c$, which is a robust and easy-to-calculate parameter of the model, being only a multiplicative factor in front of the normalized elongation $\xi$. Even so, the differences in the estimations of $L_c$, even if small among the different functions, are still appreciable, as reported in Table~\ref{table4-4m}. One may think that the estimation of the contour length is a natural outcome of a linear fit at high forces. However in Table~\ref{table4-4m} it can be seen that the best prediction for $L_c$ is found by using the na\"ive formula $\xi_L$, almost independently of the $k$ values. Table~\ref{table3-4m} shows the results of the three parameters fit performed by fixing in the value of $L_c$ as calculated by $\xi_L$. This way the prediction of the remaining parameters strongly improves for almost all the functions, leaving $\xi_{EW_2}$ as the most performant function, which overall gives the best outcomes for both intermediate and low elastic constant values.

\begin{table}[tbp]
\begin{center}
\begin{tabular}{c c c c c c | c c }
\hline \hline
                       &             & $\xi_{EW_0}$   & $\xi_{EW_1}$   &  $\xi_{EW_2}$   &$\xi_{L}$    & $\xi_{EF}$      & $L_c(1\!+\!f/S)$  \\
                       &             &  $f\!>\!30$            &   $]0,50]$   &   $]0,50]$    &  $f\!>\!30$    &  $f\!>\!30$          &  $f\!>\!30$       \\
\hline
                         & $k$        &  7.4                &  {\bf 10.08}     &  {\bf 9.89}               &  1.73         &  1.70            & ({\bf 10.00})  \\ 
$k=10$             & $k_b$    &  5.14              &  12.75             &  {\bf 9.48}       &  30.64       &  -                  &  -    \\
                         & $l_0$     &  1.359            &  0.972             & {\bf 1.002}      &  5.837       &  5.902          &  -      \\
                         & $L_c$    &  1.007$^*$     &  1.007$^*$     & 1.007$^*$       & {\bf 1.007}     &  1.003      & {\bf 1.000}  \\
\hline
                          & $k$        & 114.9          & 125.2               & {\bf 109.9}      &  58.63        & 43.96        &  ({\bf 95.40})    \\
$k=100$           & $k_b$     & 4.50             & 11.43               & {\bf 9.87}        &  7.97          & -                &  -   \\
                         & $l_0$      & 0.882           & 0.787               & {\bf 0.910}      & 1.709         & 2.296       &  -   \\
                         & $L_c$     & 1.002$^*$    & 1.002$^*$        & 1.002$^*$      & {\bf 1.002}  &  0.998       &  0.979 \\

\hline
                            & $k$         & 1165              & 898.1          & 925.2          & {\bf 946.6}     & 386.2        & (725.0)\\
$k\!=\!1000$            & $k_b$     & 11.33             & 8.15            & 8.32            & {\bf 9.47}       &  -               &  -       \\
                            & $l_0$      & 0.866             & 1.135          & 1.110          & {\bf 1.055}     &  2.396       &  -         \\
                            & $L_c$     & 1.001$^*$      & 1.001$^*$   & 1.001$^*$  & {\bf 1.001}     &  0.991       &  0.970 \\

\end{tabular}
\end{center}
\caption{Three parameters fit ($k$, $k_b$, $l_0$) by using the analytic formulas. The asterix indicates that the value of $L_c$ is fixed to the outcome of the na\"ive formula  $\xi_L$ (Eq.~(\ref{xi_L})). The bolded values indicate the best fit outcomes.}
\label{table3-4m}
\end{table}

\section{Summary and comments.}
This paper presents an analytical derivation of the partition function of the extensible discrete WLC polymer model. The difficulty of the model guided us to find two different approaches. The first one in solving numerically the problem by means of a Transfer Matrix Evaluation which has been double checked with Langevin simulations. The second one, by writing down some approximated handy formulas that allows the fit estimation of the parameters of the model with increased fidelity with respect to the past.

As known, the partition function of the discrete extensible WLC model under stretching is not factorable in single bonds. Nevertheless, with the limit of the numerical evaluation, the TM gives the exact outcomes, calculated both at the thermodynamic limit as well as at finite polymer length.

Analogously as in the previously calculated extensible FJC (Ref.~\cite{2019AF-FF}), the analytic formulas of the end-to-end distance here obtained are a combination of elementary functions easy to implement in any fit of experimental data of semi-flexible polymers. The fit obtained with the analytic approximations have confirmed the Rosa equation with the na\"ive link extension as giving the best parameter predictions at high forces, while the new formula $\xi_{EW_2}$ obtains the best predictions at low forces, especially if the number of free parameters can be reduced to three. 

Overall, the derived functions represent the most approximated expressions to the analytic discrete extensible WLC model found at the date. Hopefully these functions will help to improve the determination of the polymer magnitudes through a new analysis of experimental measures as described in this work.

\vspace{0.5cm}
\emph{Acknowledgments.---}
The authors acknowledge the Grant PID2020-113582GB-I00 funded by MCIN/AEI/ 10.13039/501100011033, and the support of the Aragon Government to the Recognized group `E36\_20R F\'isica Estad\'istica y no-lineal (FENOL)'. AF also acknowledges the funds of the European Union-NextGenerationEU, and the Spanish Ministerio de Universidades through the grant BOA 139 (31185) 01/07/2021. The authors also thank Dr J. L. Garc\'ia-Palacios and Prof L. M. Flor\'ia for the very useful discussions on the subject.

\end{document}